\documentclass[journal=jpclcd,manuscript=letter]{achemso}

\usepackage{graphicx}
\usepackage{float}
\usepackage{color}
\usepackage{bigints}
\usepackage[normalem]{ulem}
\usepackage{bm}
\usepackage{amsmath,amsfonts,amssymb} 
\usepackage{latexsym}

\newcommand{\onlinecite}[1]{\hspace{-1 ex} \nocite{#1}\citenum{#1}}

\setkeys{acs}{keywords = true}

\title{Cavity-modified exciton dynamics in photosynthetic units}

\author{Roc\'io S\'aez-Bl\'azquez}

\affiliation{Departamento de F\'isica Te\'orica de la Materia
Condensada and Condensed Matter Physics Center (IFIMAC),
Universidad Aut\'onoma de Madrid, E- 28049 Madrid, Spain}

\author{Johannes Feist}

\affiliation{Departamento de F\'isica Te\'orica de la Materia
Condensada and Condensed Matter Physics Center (IFIMAC),
Universidad Aut\'onoma de Madrid, E- 28049 Madrid, Spain}

\author{Elisabet Romero}

\affiliation{Institute of Chemical Research of Catalonia (ICIQ),
The Barcelona Institute of Science and Technology (BIST), Tarragona,
Spain}

\author{Antonio I. Fern\'andez-Dom\'inguez}

\affiliation{Departamento de F\'isica Te\'orica de la Materia
Condensada and Condensed Matter Physics Center (IFIMAC),
Universidad Aut\'onoma de Madrid, E- 28049 Madrid, Spain}
\email{a.fernandez-dominguez@uam.es}

\author{Francisco J. Garc\'ia-Vidal}

\affiliation{Departamento de F\'isica Te\'orica de la Materia
Condensada and Condensed Matter Physics Center (IFIMAC),
Universidad Aut\'onoma de Madrid, E- 28049 Madrid, Spain}
\altaffiliation{Donostia International Physics Center (DIPC),
E-20018 Donostia/San Sebasti\'an, Spain} \email{fj.garcia@uam.es}

\begin{tocentry}
\begin{center}
\includegraphics[angle=0,width=1\textwidth]{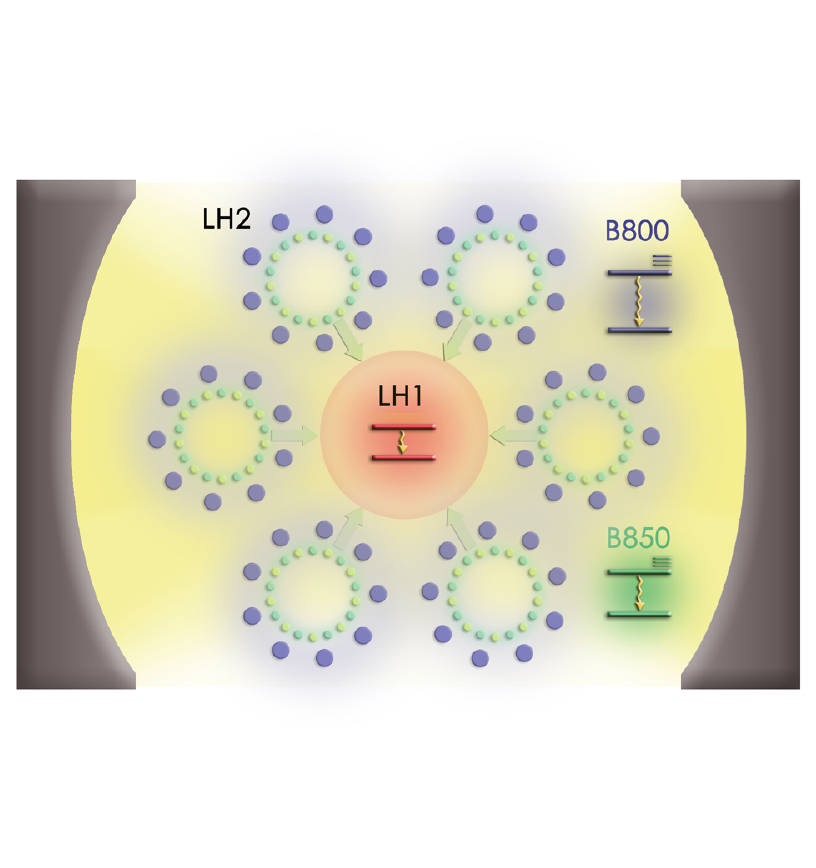}
\end{center}
\end{tocentry}


\begin{document}

\begin{abstract}
Recently, exciton-photon strong coupling has been proposed as a
means to control and enhance energy transfer in ensembles of
organic molecules. Here, we demonstrate that the exciton dynamics
in an archetypal purple bacterial photosynthetic unit, composed of
six LH2 antennas surrounding a single LH1 complex, is greatly
modified by its interaction with an optical cavity. We develop a
Bloch-Redfield master equation approach that accounts for the
interplay between the B800 and B850 bacteriochlorophyll molecules
within each LH2 antenna, as well as their interactions with the
central LH1 complex. Using a realistic parametrization of both
photosynthetic unit and optical cavity, we investigate the
formation of polaritons in the system, revealing that these can be
tuned to accelerate its exciton dynamics by three orders of
magnitude. This yields a significant occupation of the LH1
complex, the stage immediately prior to the reaction center, with
only a few-femtosecond delay after the initial excitation of the
LH2 B800 pigments. Our theoretical findings unveil polaritonic
phenomena as a promising route for the characterization, tailoring,
and optimization of light-harvesting mechanisms in natural and
artificial photosynthetic processes.
\end{abstract}

\maketitle

Light-harvesting (LH) complexes play a crucial role in the process
of photosynthesis~\cite{Scholes2011,Croce2014}. They are
responsible for collecting, retaining, and funnelling solar
energy~\cite{McConnell2010,Mirkovic2017} into the reaction
centers, where its conversion into chemical energy takes
place~\cite{Romero2017}. These pigment-protein compounds absorb
the incident photons and convey the resulting electron-hole
excitations through F\"orster-like, dipole-dipole interactions
between neighboring molecules~\cite{Novoderezhkin2010,Ye2012}.
This mechanism is slower than vibrational dephasing in the system,
which makes the transport process effectively
incoherent~\cite{Cheng2009,Fassioli2014}. Moreover, thanks to the
extremely slow non-radiative decay inherent to bacteriochlorophyll
molecules, energy transfer in photosynthetic membranes can range
micrometric distances and take nanosecond times while having
efficiencies approaching $100\%$~\cite{Caycedo2010,Timpmann2014}.
A paradigmatic example of phototrophic organisms, widely studied
in the literature, are purple
bacteria~\cite{Hunterbook,Codgell2006} such as
\emph{Rhodopseudomonas acidophila}, in whose photosynthetic
membranes two different complexes can be
identified~\cite{Nagarajan1997}: LH2, which act mainly as optical
antennas, and LH1, which deliver the excitation to the reaction
center they enclose. Although the arrangement and distribution of
both complexes within the bacterial membrane depends on the
ambient and light intensity conditions, there is usually a number
of LH2 in the vicinity of every LH1 and attached reaction
center~\cite{Caycedo2010,Cleary2013}.

In recent years, much research attention has been focused
on exploring the opportunities that the phenomenon of
exciton-photon collective strong coupling~\cite{Lidsey1998} brings
into material science~\cite{Ebbesen2016}. The coupling between an
excitonic platform and the electromagnetic modes supported by an
optical cavity gives rise to polaritons, hybrid states whose
formation requires that the interaction between light and matter
become faster than their respective decay channels. Experimental
and theoretical studies demonstrate that the appropriate tailoring
of polaritonic characteristics in organic semiconductors and
ensembles of organic molecules can yield a large enhancement of
the efficiency and spatial range of charge and exciton
conductance~\cite{Orgiu2015,Feist2015,Schachenmayer2015,Schlawin2019} and energy
transfer~\cite{Coles2014,Zhong2016,Zhong2017,Du2018,SaezBlazquez2018}
in these systems. The coherent and delocalized nature of
polaritons plays a crucial role in these phenomena. On the one
hand, it allows energy transfer within a time scale set by the
so-called Rabi frequency (collective coupling
strength)~\cite{GlezBallestero2015}. On the other hand, it makes
the process nonlocal and robust to disorder within a length scale
comparable to the optical wavelength~\cite{GarciaVidal2017}.

It has been recently shown that plasmonic nanostructures can modify
the optical properties of LH2
antennas~\cite{Wientjes2014,Wientjes2016,Tsargorodska2016,Caprasecca2018}.
Moreover, experimental evidence of collective strong coupling in
ensembles of living bacteria has been
reported~\cite{ColesSmith2014}, giving rise even to the concept of
\textit{living polaritons}~\cite{ColesLidzey2017}. In Ref.~\onlinecite{ColesSmith2014} 
a Rabi splitting of around $150$ meV has been reported, 
implying that about one thousand chlorosomes present in 
green sulfur bacteria are coherently coupled to a cavity photon. 
In the absence of a cavity, the study of exciton transport in photosynthetic materials 
has been triggered by the prospect of transferring this 
knowledge to human-made energy-harvesting structures. 
In this Letter, we go a step further by assesing the impact that the interaction with an
optical cavity has on the efficiency of exciton transport taking place in purple
bacterial photosynthetic units (PSUs) formed by several LH
complexes. Using Bloch-Redfield
theory~\cite{Bloch1957,Redfield1957}, which allows us to describe
vibration-assisted incoherent interactions among
bacteriochlorophyll
pigments~\cite{Novoderezhkin2010,Fassioli2014}, we construct first
a quantum master equation describing a single LH2 antenna,
involving 27 interacting pigments of three different families
(B800, B850a and B850b). Our model reproduces experimental
absorption spectra of free-standing LH2. Next, we consider an
archetypal PSU configuration~\cite{Caycedo2010,Ritz2001}: a ring
of six LH2 antennas surrounding a single LH1 complex. We extend
our master equation to the whole PSU, including incoherent
interactions among neighboring pigments within different LH
complexes. By introducing pigment-photon coupling terms in the
Hamiltonian, we study the formation of polaritons in the system,
with special emphasis on the cavity characteristics. We find that
strong coupling in realistic cavities can accelerate PSU exciton
dynamics by a factor $\sim10^3$, which leads to a considerable
population of the LH1 complex within only a few femtoseconds after
the initial excitation of LH2 B800 pigments. Our model also
reveals how the contribution of the different polaritonic states
to this fast population transfer depends on the frequency of the
cavity mode and its effective volume (or pigment-photon coupling
strength).

\begin{figure}[!t]
\includegraphics[width=0.7\linewidth]{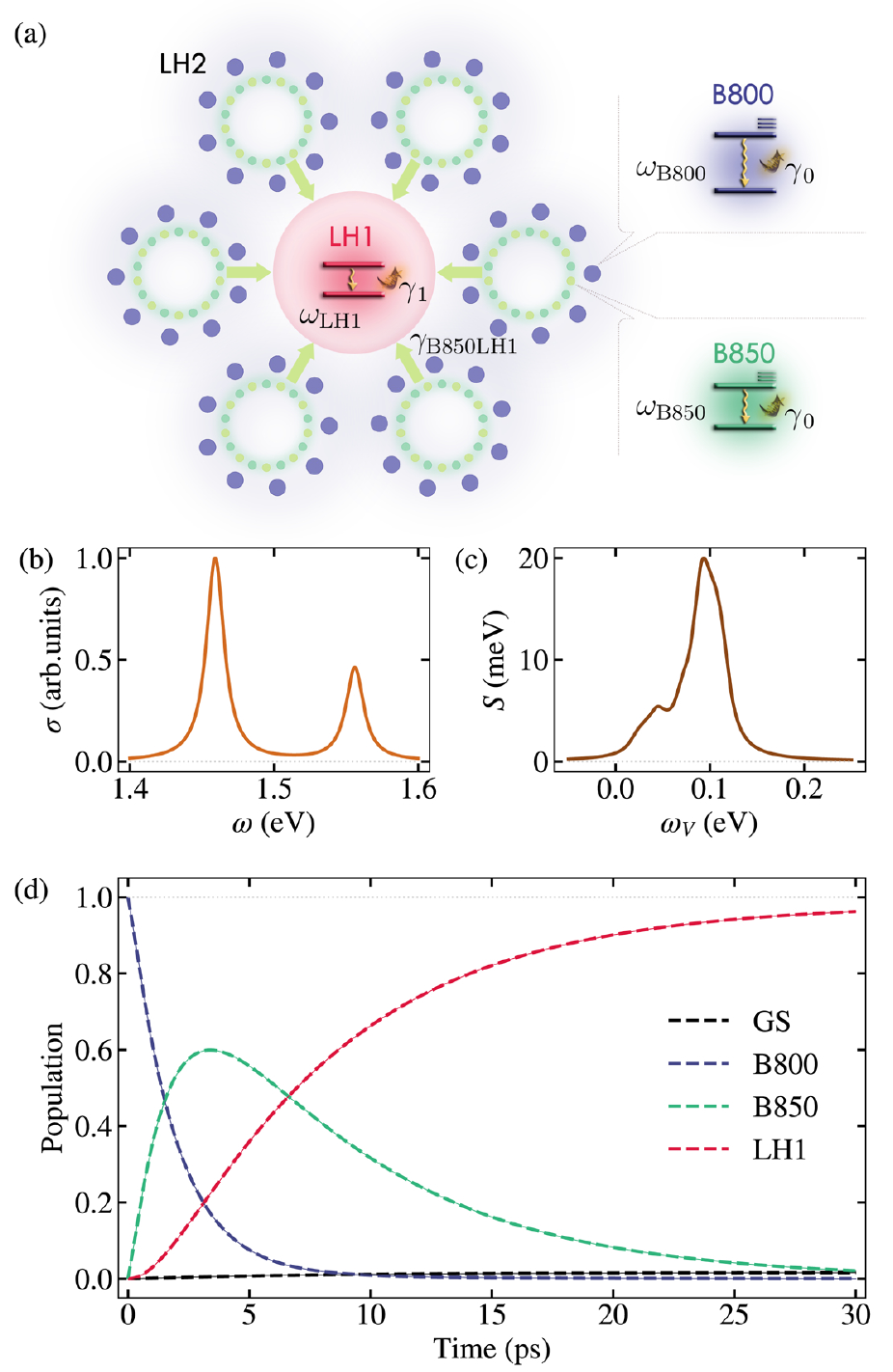}
\caption{(a) Sketch of the PSU considered in this work: 6 LH2
antennas, comprising 9 B800 and 18 B850 molecules each, surrounding a
single LH1 complex. The insets show the two-level system exciton model of
B-molecules and LH1, which experience both radiative and
vibrational decay. (b) Absorption spectrum of a single LH2
complex, including disorder and inhomogeneous broadening. (c)
Vibrational spectral density, $S(\omega)$, for all LH2 pigments.
(d) Exciton population dynamics for the PSU in panel (a) in an
initial state given by the superposition of excited B800 molecules
in the six LH2 antennas.} \label{fig:1}
\end{figure}

Figure~\ref{fig:1}(a) sketches the PSU configuration under study:
six LH2 antennas arranged around a single LH1 complex. The latter
is formed by a number of B875 pigments~\cite{Mirkovic2017}, which
are the final stage of the exciton transfer mechanism we analyze
here. Taking this into account, we use a simplified model for the
LH1 complex (red circle), valid in the low population regime. We
treat it as a single two-level system with transition frequency
$\omega_{\rm LH1}=1.417$ eV~\cite{Caycedo2017}. Our attention is
focused on the LH2 antennas, which we describe in more detail.
They are composed of $N_{\rm LH2}=27$ pigments, distributed in a
double-ring
structure~\cite{McDermott1995,Cupellini2016,Caycedo2018}: while
nine B800 molecules (blue dots) form one of the rings, the other
comprises nine pigment dimers made up of a B850a and a B850b
molecule each (light and dark green dots). In our model, the
pigments in each LH2 complex are modelled as interacting two-level
systems leading to the Hamiltonian
\begin{equation}
\hat{H}_{\rm LH2}=\sum_{i=1}^{N_{\rm
LH2}}\omega_i\hat{\sigma}_i^\dagger\hat{\sigma}_i+\sum_{i,j=1}^{N_{\rm
LH2}} V_{ij}\{\hat{\sigma}_i^\dagger\hat{\sigma}_j +
\hat{\sigma}_j^\dagger\hat{\sigma}_i\}, \label{eq:H}
\end{equation}
where $\hat{\sigma}^\dagger_i$ and $\hat{\sigma}_i$ are the
creation and annihilation operators of molecular excitations, and $\omega_i$ their corresponding energies. The
parameters in Eq.~\eqref{eq:H} are taken from
Refs.~\onlinecite{Sauer1996}~and~\onlinecite{Tetriak2000}. The
freestanding-pigment energies are set to $\omega_{\rm B800}=1.549$
eV and $\omega_{\rm B850a,b}=1.520$ eV. The intra-ring
nearest-neighbor (second-nearest-neighbor) couplings are $V_{\rm
B800B800}=3$ meV, $V_{\rm B850aB850a}=6$ meV,
$V_{\rm B850bB850b}=4$ meV,  and $V_{\rm B850aB850b}=-33(-36)$ meV. Finally,
the inter-ring nearest-neighbor interaction is parameterized by
$V_{\rm B800B850a}=-3$ meV and $V_{\rm B800B850b}=-1$ meV.

The diagonalization of the Hamiltonian above yields the exciton
energies of the LH2 complex. The small inter-ring couplings above
translates into excitons strongly localized at B800 or B850(a,b)
pigments, with little hybridization among
them~\cite{Cupellini2016,Caycedo2018}. We check the validity of
our model by calculating the absorption spectrum for a single,
isolated LH2 antenna. We first compute the transition matrix element of
the total dipole moment operator $\hat{M}=\sum_{i=1}^{N_{\rm
LH2}}\mu_i\hat{\sigma}_i^\dagger$ ($\mu_{\rm B800}=\mu_{\rm
B850}=6.13$ D~\cite{Linnanto1999}) for the excitonic eigenstates of
Eq.~\eqref{eq:H}. The absorption spectrum is built as a sum of
Lorentzian contributions centered at the excitonic energies and
weighted by the square of the corresponding matrix element of
the dipole moment operator. Their width was set to 15
meV for B800 and B850 excitons, in order to take
into account the disorder and inhomogeneous broadening inherent to
the measurements performed on ensembles of LH2
complexes~\cite{Oijen1999}. The spectrum obtained this way is
rendered in Fig.~\ref{fig:1}(b), which reproduces the
double-peaked absorption profile reported
experimentally~\cite{Oijen1999,Hildner2013}, with maxima around 1.44
eV (860 nm) and 1.55 eV (800 nm).

We use a Bloch-Redfield master equation~\cite{Novoderezhkin2010}
to describe the vibrational dissipation and incoherent
interactions experienced by B800 and B850 molecules. This requires
the inclusion of the vibronic spectral density of the pigments,
$S(\omega)$. Fig.~\ref{fig:1}(c) plots $S(\omega)$ in our
calculations (the same for B800 and B850 molecules), parameterized
using the Franck-Condon factors in Ref.~\onlinecite{Zazubovic2001}
and a thermal line broadening in agreement with
Ref.~\onlinecite{Caro1994}. Lindblad terms of the form
$\tfrac{\gamma_0}{2}\mathcal{L}_{\hat{\sigma}_i}[\hat{\rho}]=
\tfrac{\gamma_0}{2}[2\hat{\sigma}_i\hat{\rho}\hat{\sigma}_i^\dagger-\{\rho\hat{\sigma}_i^\dagger\hat{\sigma}_i\}]$,
acting on all pigment annihilation operators are also included in
the master equation, weighted by a decay rate $\gamma_0=1$
$\mu$eV, which reflects the $\sim 1$ ns lifetime of all the
molecules in the LH2 complex~\cite{Mirkovic2017}. Note that the
exciton widths introduced in the cross section calculations are
$\sim10^4$ times larger than this value.

The master equation for the whole PSU is built next. It is
composed by blocks, corresponding to the six LH2 antennas and the
central LH1 complex, only connected through Lindblad terms of the
form $\tfrac{\gamma_{\rm
B850LH1}}{2}\mathcal{L}_{\hat{\sigma}^\dagger_{\rm
LH1}\hat{\sigma}_{n,1}}[\hat{\rho}]$. These act on the product of
the annihilation operator for the B850 molecule (which we label as
$i=1$) in the n-th LH2 antenna that is located next to the LH1
complex, and the LH1 creation operator. The associated decay rate
is set to $\gamma_{\rm B850LH1}=2$ meV, which yields LH2-LH1
transition rates in agreement with experiments~\cite{Ritz2001}.
This is shown in Fig.~\ref{fig:1}(d), which plots the population
transients for a freestanding PSU. The initial state corresponds
to the coherent superposition of excitations in all the B800
pigments, which mimics an experimental setup in which the PSU is
pumped by an ultrashort laser pulse centered around $800$ nm. We
can observe that this state (blue line) decays within $\sim 3$ ps,
feeding population into the LH2 B850 molecules (green line). These
in turn carry the excitation to the LH1 complex (red line), whose
population grows within a $\sim 20$ ps time scale after the
initial excitation. The time interval in Fig.~\ref{fig:1}(d) is
much shorter than $\gamma_0^{-1}$, and the ground state (shown in
black line) is negligibly populated in the whole exciton transfer
process. 
Note that we have taken $\gamma_1=0$, see
Fig.~\ref{fig:1}(a), to avoid the decay of the LH1 excitations
into the ground state.

\begin{figure}[!t]
\includegraphics[width=0.7\linewidth]{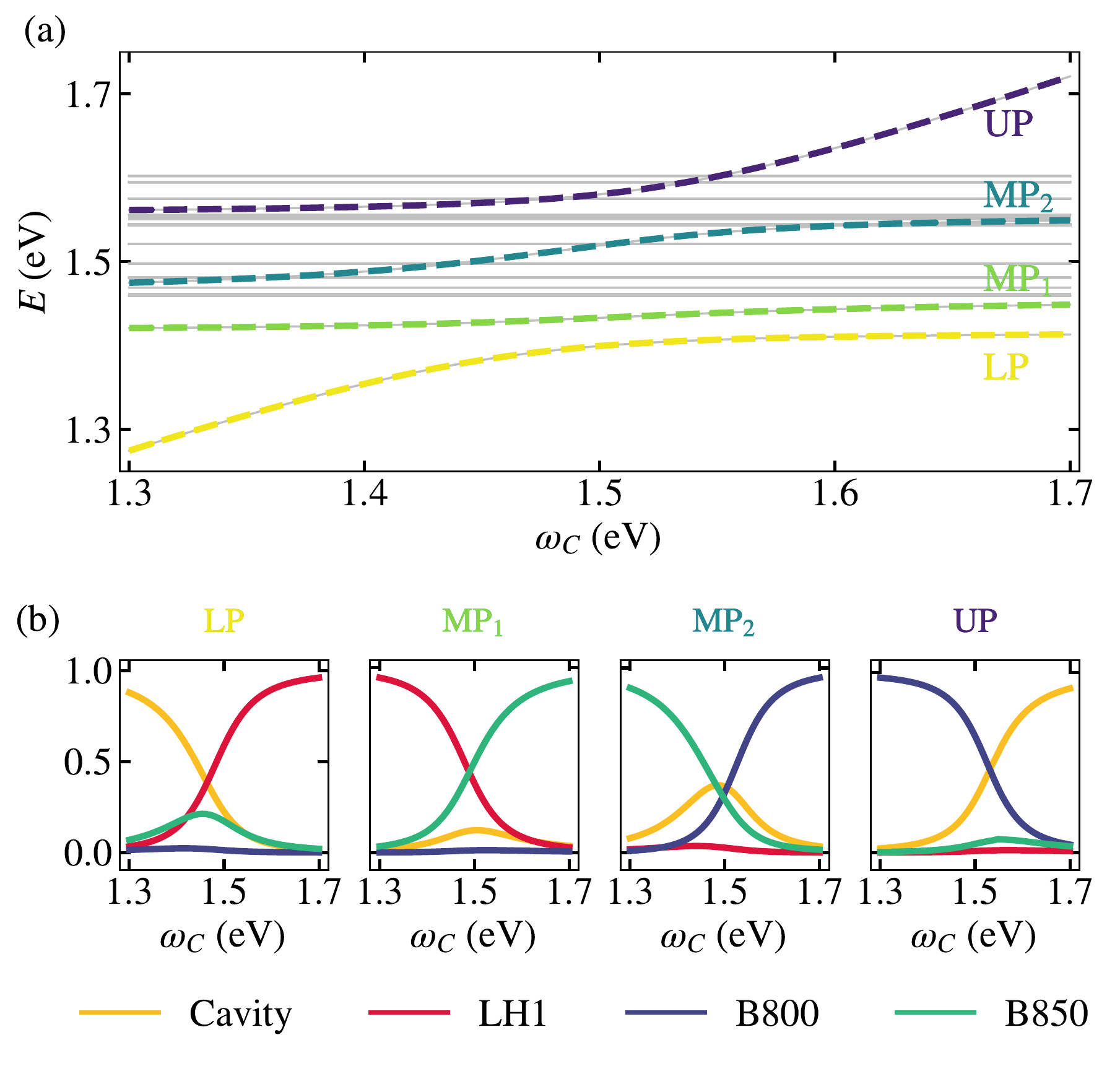}
\caption{(a) Energies of the lower (LP, yellow), middle (MP$_1$,
green, and MP$_2$, blue), and upper (UP, violet) polaritons versus
the cavity frequency, $\omega_{C}$, and for $g_0=5$ meV. (b)
Coefficients representing the cavity (yellow), LH1 (red), B800
(blue) and B850 (green) content of the four polaritons in panel
(a) as a function of $\omega_{C}$.} \label{fig:2}
\end{figure}

In order to analyze the effect of strong coupling in the PSU
exciton dynamics, we add new terms to the freestanding PSU
Hamiltonian, describing the coherent interactions between pigments
and cavity photons,
\begin{eqnarray}
\hat{H}&=&\omega_C\hat{a}^\dagger\hat{a}+\omega_{\rm
LH1}\hat{\sigma}_{\rm LH1}^\dagger\hat{\sigma}_{\rm
LH1}+\sum_{n=1}^6\hat{H}_{{\rm LH2},n}+ \notag
\\
&&+g_0\left[\sum_{n=1}^6\sum_{i=1}^{N_{\rm
LH2}}(\hat{\sigma}_{n,i}^\dagger\hat{a}+\hat{\sigma}_{n,i}\hat{a}^\dagger)+\eta(\hat{\sigma}_{\rm
LH1}^\dagger\hat{a}+\hat{\sigma}_{\rm LH1}\hat{a}^\dagger)\right],
\label{eq:H2}
\end{eqnarray}
where $\omega_C$ is the cavity frequency, $\hat{a}^\dagger$ and
$\hat{a}$ are the creation and annihilation operators for the
cavity mode, $g_0$ is the photon coupling strength to all LH2
pigments, and $\eta=3.8$ is the dipole moment of the LH1 complex
normalized to $\mu_{\rm B800}$~\cite{Mirkovic2017}. There,
$\hat{H}_{{\rm LH2},n}$ is the Hamiltonian for the n-th LH2
antenna, see Eq.\eqref{eq:H}. As a result of the interaction
between molecules and cavity and the symmetry of the structure, 
only four hybrid light-matter states
arise when entering into the strong coupling regime in addition 
to the set of the so-called dark states. These states are linear combinations 
of electronic excitations within the LH1 and LH2 complexes that do not couple to the cavity mode.
Figure~\ref{fig:2}(a) renders the four polaritonic frequencies, as
well as the corresponding to the set of dark states, versus the
cavity frequency. They are obtained directly from the
diagonalization of Eq.~\eqref{eq:H2}. Color lines plot the
dispersion of the lower (LP, yellow), middle (MP$_1$ and MP$_2$,
green and blue) and upper (UP, violet) polariton bands. Note that,
certainly, the PSU dark states (grey lines) remain uncoupled to
the cavity field. In our calculations, we have taken $g_0=5$ meV.
Using that $g_0=\mu_{\rm B800}\sqrt{\omega_C/2 \epsilon_0V_{\rm
eff}}$~\cite{Giannini2011}, this value corresponds to
$V_{\rm eff}=(20 {\rm nm})^3$ at $\omega_C=1.6$ eV. This is
attainable not only in plasmonic, but also in state-of-the-art
dielectric cavities~\cite{Hu2018}.

Figure~\ref{fig:2}(b) shows, from left to right, the square of the
Hopfield coefficients for the LP, MP$_1$, MP$_2$ and UP as a
function of the cavity frequency. Calculated as $|\langle i
|\alpha \rangle|^2$, where $\alpha$ ($i$) labels the polaritonic
(exciton and cavity) states, they weight the cavity (yellow), LH1
(red), B800 (blue) and B850 (green) contents of each polariton. We
can observe that the polariton character can be greatly modified
through $\omega_C$. Note that only the LP and UP present a
substantial cavity content, but they do it at low and high cavity
frequencies, respectively. Far from these spectral regions, LP
(UP) virtually overlaps with the LH1 (B800) states. On the
contrary, the MPs present a moderate cavity component, but combine
excitonic contents corresponding to all PSU pigments. We
anticipate that the hybrid character of these states (especially
evident for MP$_2$ at $\omega_C\simeq1.5$ eV, where B800, B850 and
cavity coefficients become similar) will play a fundamental role
in the polariton-assisted population transfer in the
PSU~\cite{SaezBlazquez2018}.

Having studied the formation of polaritons in the hybrid
cavity-PSU system, and the tuning of their characteristics through the
cavity frequency, we investigate next its population
dynamics. To do so, we extend the master equation for the
freestanding PSU by the inclusion of the Hamiltonian in
Eq.~\eqref{eq:H2}, and by adding a Lindblad term describing the
cavity losses. We set the cavity decay rate to
$\gamma_C=13\,\mu$eV, which corresponds to a lifetime of 50 ps and
a quality factor $Q=\omega_C/2\gamma_C\simeq6\cdot10^4$,
parameters similar to those recently reported in deeply
subwavelength dielectric cavities~\cite{Hu2018}. Similarly to
Fig.~\ref{fig:1}(d), we set the initial state as the coherent
superposition of equally excited B800 molecules, and choose
$\omega_C=1.6$ eV and $g_0=15$ meV. Fig.~\ref{fig:2}(b) shows that
the LP is composed of LH1 excitations mostly at this cavity
frequency, which allows us to set the final stage of the
polariton-assisted energy transfer mechanism at the LH1 complex.


\begin{figure}[!t]
\includegraphics[width=0.7\linewidth]{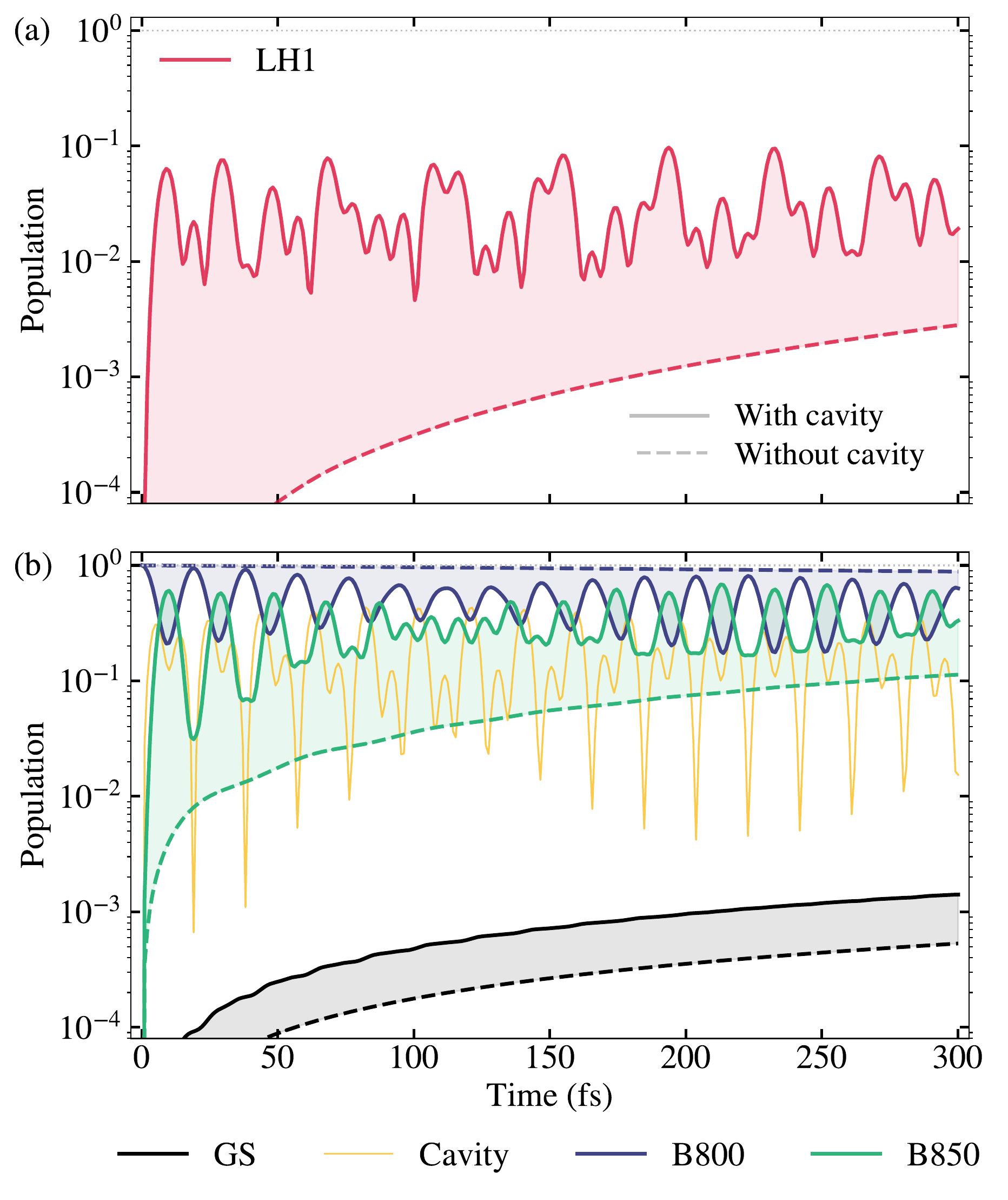}
\caption{Exciton dynamics for the PSU in Fig.~\ref{fig:1}. (a) LH1
population versus time for the PSU isolated (dashed line) and
coupled to an optical cavity with $\omega_C=1.6$ eV and $g_0=15$
meV (solid line). (b) Temporal evolution of the ground state
(black), B800 (blue) and B850 (green) populations with (solid
line) and without (dashed line) cavity. The cavity population is
shown in yellow solid line. The difference between populations
have been shaded in all cases to highlight the effect of strong
coupling.} \label{fig:3}
\end{figure}

Figure~\ref{fig:3} displays the comparison between the population
dynamics for the PSU in isolation (dashed lines) and interacting
with the cavity described above (solid lines). Fig.~\ref{fig:3}(a)
shows that exciton-photon strong coupling in the PSU gives rise to
an extremely fast occupation of the LH1 complex, which acquires a
significant population ($\sim10\%$) within only a 20 fs delay.
Note that, in absence of the cavity, the LH1 population is
negligible in this time scale, and becomes comparable only after a
few ps, see Fig.~\ref{fig:1}(d). This is the main result in this
Letter, the polariton-assisted reduction in population
transfer times taking place in PSUs by three orders of magnitude. 
Fig.~\ref{fig:3}(b) plots the
population transients for B800 (blue), and B850 (green) excitons,
both exhibiting more regular Rabi oscillations than the LH1. These
are especially apparent in the B800 case. 
Its occupation remains constant and
close to unity for the freestanding PSU, but the occurrence of
strong coupling gives rise to a coherent energy exchange that
feeds population into the other excitonic and cavity (yellow line)
states. Black lines correspond to the ground state, whose
population is larger in the strong coupling regime. This is a
consequence of the short lifetime of the cavity relative to the
PSU pigments ($\gamma_C\sim\gamma_0/20$). This loss channel can be
mitigated by using nanocavities with higher quality factors.

\begin{figure}[!t]
\includegraphics[angle=0,width=0.7\linewidth]{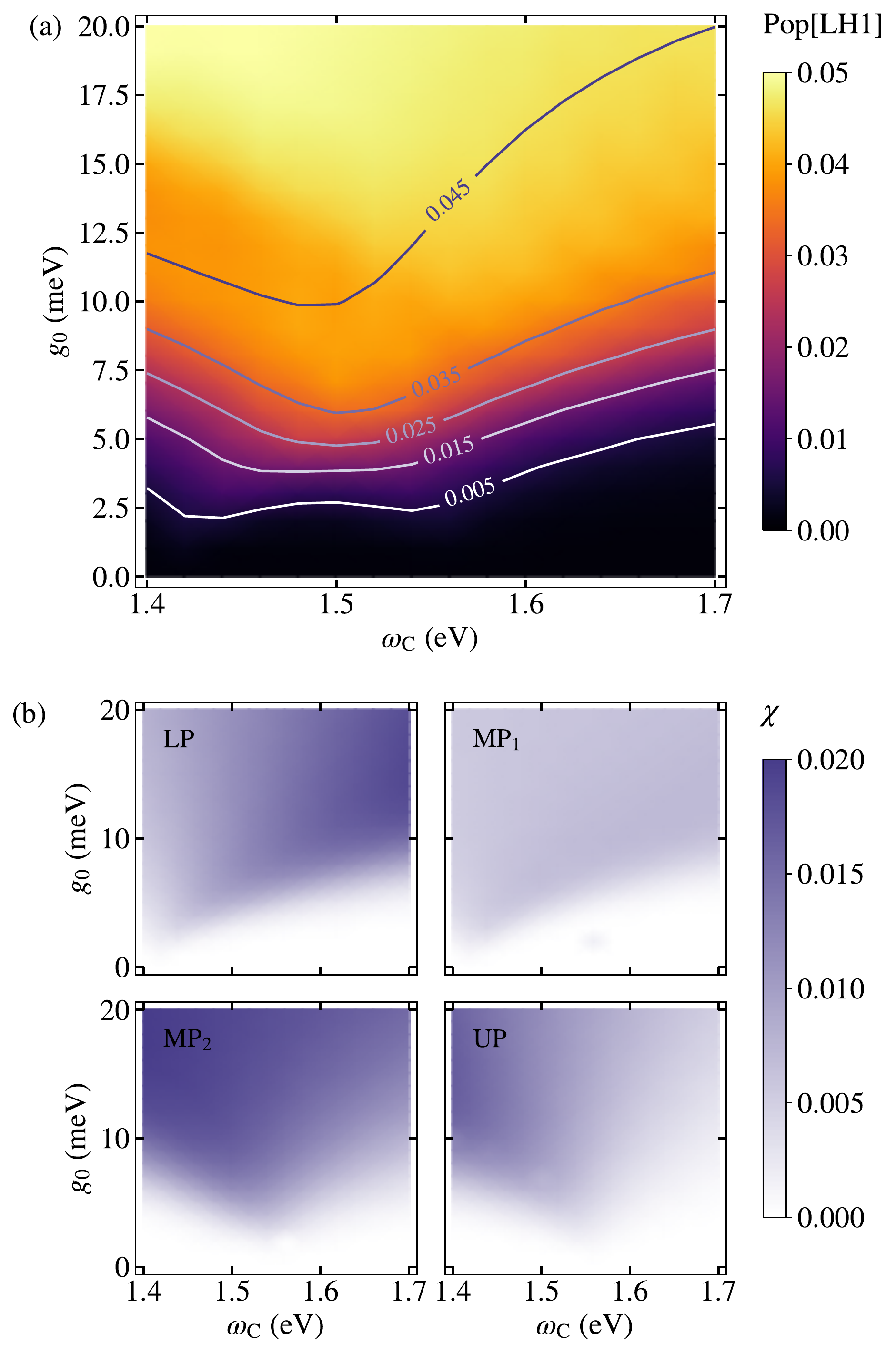}
\caption{(a) LH1 population averaged over the first 300 fs after
the excitation of the LH2 B800 pigments versus cavity frequency
and photon-exciton coupling strength. Color solid lines render
contours of the magnitude $\sum_\alpha\chi_\alpha$ (see panel
below). (b) Polariton component of B800 and LH1 excitations,
$\chi_\alpha= | \langle \alpha|{\rm
LH1}\rangle |^2  |\langle \alpha  | {\rm B800 } \rangle |^2$, with
$\alpha={\rm LP, MP_1, MP_2}$ and ${\rm UP}$.\label{fig:4}}
\end{figure}

Up to this point, we have demonstrated that the exciton dynamics
in PSUs is greatly modified due to the interaction with a
particular optical cavity configuration. We have also shown that
this phenomenon is mediated by the polaritons that emerge in the
system, whose character varies strongly with the frequency of the
cavity. In the following, we shed insights into both findings by
investigating the dependence of the B800-to-LH1 population
transfer and the polaritonic content of B800 and LH1 excitations on the two
parameters set by the optical cavity: $\omega_C$ and $g_0$.
Figure~\ref{fig:4}(a) displays a contourplot of the LH1 population
averaged over the first 300 fs after the excitation of the B800
molecules (the time span in Fig.~\ref{fig:3}). Note that the
temporal averaging naturally removes peak effects related to the
irregular Rabi oscillations apparent in Fig.~\ref{fig:3}(a). We
can observe that by increasing the exciton-photon coupling
(reducing the cavity mode volume) the LH1 population grows, with a
much weaker dependence on $\omega_C$. This is a surprising result,
given the strong dependence of the Hopfield coefficients, and
therefore the polariton character, on the cavity frequency shown
in Fig.~\ref{fig:2}(b).

Figure~\ref{fig:4}(b) plots the combination of content of B800 and LH1 excitations on the
various polaritons $\alpha$, $\chi_\alpha= | \langle \alpha|{\rm
LH1}\rangle |^2  |\langle \alpha  | {\rm B800 } \rangle |^2$, 
as a function of the cavity frequency and
coupling strength. We can observe that, as expected, $\chi_\alpha$
grows with $g_0$ in all cases, as the light-matter hybridization
increases with this parameter. However, the B800 and LH1 projections over
the different polaritons varies much with $\omega_C$. For large
couplings ($g_0>10$ meV), $\chi_{\rm LP}$ dominates LH1 excitations
for blue-detuned cavities, whereas $\chi_{\rm MP_2}$ and
$\chi_{\rm UP}$ are the largest contributions for red-detuned
ones. On the other hand, for modest couplings ($g_0\lesssim10$ meV), $\chi_{\rm LP}$
and $\chi_{\rm MP_1}$  are largest for red-detuned cavities, while
$\chi_{\rm MP_2}$ and $\chi_{\rm UP}$ present a maximum within the
spectral window between 1.5 and 1.6 eV. These panels indicate that
the interplay among the different polaritons play a crucial role
in the fast B800-to-LH1 population transfer in PSUs. 
This conclusion is
supported by the contour lines in Fig.~\ref{fig:4}(a), which render 
$\sum_\alpha\chi_\alpha$, showing that this magnitude presents the
same dependence on $\omega_C$ and $g_0$ as the LH1 averaged
population.

To conclude, we have investigated exciton-photon strong coupling
in an archetypal purple bacterial photosynthetic unit, comprising
six LH2 antennas surrounding a single LH1 complex. We have
developed a master equation combining Bloch-Redifield and Lindblad
approaches to describe the vibration-assisted incoherent
interactions among the B800, B850 and LH1 excitons, as well as
their coherent coupling to the electromagnetic mode supported by
an optical cavity. Using this tool, we have explored the formation
of polaritons in the system, analyzing their dependence on the
cavity configuration. We have revealed that these hybrid
light-matter states yield a three orders of magnitude reduction in the
B800-to-LH1 population transfer times, leading to a significant LH1
occupation in a few-femtosecond time scale. We believe that our
theoretical findings demonstrate the potential of exciton-photon
strong coupling not only for the characterization of
light-harvesting phenomena in natural photosynthesis, but also as
a means for the design and optimization of artificial
photosynthetic systems.

\section{Acknowledgments}

This work has been funded by the European Research Council under
Grant Agreements ERC-2011-AdG 290981 and ERC-2016-STG-714870, the
EU Seventh Framework Programme (FP7-PEOPLE-2013-CIG-630996), and
the Spanish MINECO under contracts MAT2014-53432-C5-5-R and
FIS2015-64951-R, and through the ``Mar\'ia de Maeztu''
programme for Units of Excellence in R\&D (MDM-2014-0377), as well as 
through two Ramon y Cajal grants (JF and AIFD). We also acknowledge support 
by the QuantERA program of the European Commission with funding by 
the Spanish AEI through project PCI2018-093145. E.R. thanks the ICIQ Foundation for the Starting Career Programme and the Generalitat de Catalunya for the CERCA Programme.

\end{document}